# Molecular Dynamics Simulations of Tensile Behaviour of Copper

G. Sainath*, V.S. Srinivasan, B.K. Choudhary, M.D. Mathew and T. Jayakumar

*Metallurgy and Materials Group*
*Indira Gandhi Centre for Atomic Research, Kalpakkam-603102, Tamil Nadu, India*
*(\*e-mail: sg@igcar.gov.in)*

**Abstract**

Molecular dynamics simulations on tensile deformation of initially defect free single crystal copper nanowire oriented in <001>{100} has been carried out at 10 K under adiabatic and isothermal loading conditions. The tensile behaviour was characterized by sharp rise in stress in elastic regime followed by sudden drop at the point of dislocation nucleation. The important finding is that the variation in dislocation density is correlated with the observed stress-strain response. Several interesting micro-structural features were observed during tensile deformation such as slip, phase transformation and pentagonal structure in necking region affecting the plastic deformation behaviour of single crystal copper nanowire.

## 1. Introduction

Metallic nanowires are of great technological importance because of their appealing properties such as high strength, excellent thermal and electrical conductivity, and quantized conductance. In recent years, mechanical properties of metallic nanowires of copper and gold have been studied extensively by using density functional theory calculations and molecular dynamics (MD) simulations [1] because of their potential applications in Nano/Macro Electro Mechanical systems (NEMS/MEMS). Although, the mechanical behaviour of metallic materials in bulk form is well understood, it can not be directly extended to nanoscale materials due to their size effects and high surface to volume ratio. Therefore, an examination of the deformation and fracture behaviour of nanoscale materials either by experimentation and/or simulation becomes very important in the present day context.

Present investigation aims at simulating the tensile behaviour of single crystal copper nanowire by MD technique under isothermal and adiabatic conditions. Suitable embedded atom method (EAM) potential for copper has been chosen for simulations. Variation of dislocation density with strain has been evaluated and correlated with stress-strain response. Interesting micro-structural features observed during plastic deformation have been discussed briefly.

## 2. Simulation Details

Molecular dynamics simulation on tensile deformation behaviour of initially defect free single crystal Cu nanowire have been carried out by using Large-scale Atomic/Molecular Massively Parallel Simulator (LAMMPS) package [2]. The simulation box has a dimensions of 15$a$x15ax38a which contains 34200 atoms, where a is lattice parameter of Cu. Free boundary conditions were imposed in x[100] and y[010] directions and periodic boundary conditions in z[001] direction (the loading direction). The simulation box with this boundary conditions mimics the pseudo infinite nanowire. Simulation box had three regions, two rigid regions of top and bottom and the remaining middle portion as the active deformation region. The initial velocities of atoms are chosen randomly from finite temperature Maxwell distribution. EAM potential due to Mishin et al [3] has been used to describe the inter-atomic forces between Cu atoms. The model system is equilibrated for 20 ps in micro-canonical (NVE) ensemble with zero applied pressure using Berendsden Barostat and to a temperature of 10 K using velocity rescaling method. Upon completion of the equilibration process, the deformation is carried out in a step wise manner with a time step of 1 fs by applying the constant strain rate of $1\times10^9$ s$^{-1}$ along z-direction and imposing stress-free condition in other two directions. The average stress is calculated using the Virial definition of stress. In order to investigate the influence of thermal loading conditions on deformation, two conditions have been considered during simulation. The first is the isothermal condition, where the temperature is not allowed to increase under the deformation and maintained at 10 K by velocity rescaling method. The second condition is adiabatic, where heat transfer is not allowed, which increases the temperature during straining.

The atomic snapshots of specimens at various stages of tensile deformation were generated using Atom-eye package [4]. A scalar quantity known as atomic centro-symmetry parameter(CSP) is used to visualize the defects and lattice rearrangements developed during deformation. The dislocations and stacking faults were viewed by using DXA package [5]. In addition, this also provides information about dislocation density, i.e., total length of dislocations per unit volume.

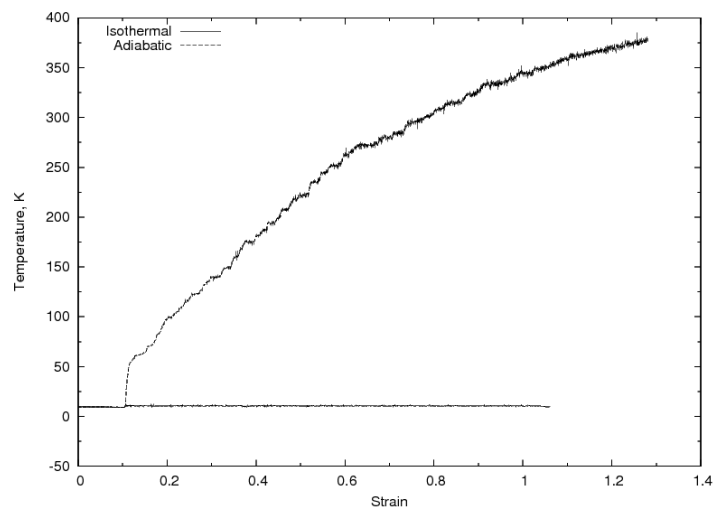

Fig. 1. Average temperature of the specimen as a function of strain under isothermal and adiabatic conditions.

## 3. Results and Discussion

### 3.1 Tensile behaviour of single crystal Cu nanowire

The tensile behaviour of Cu nanowire has been simulated under isothermal and adiabatic conditions. Under adiabatic testing conditions, temperature of the specimen increases with straining whereas it remains constant under isothermal conditions (Fig. 1). The estimated temperature is the mean temperature in the specimen at any value of strain. It can be seen from Fig. 1 that the temperature increases continuously from the onset of plastic deformation until it reaches a value of about 380 K near failure which is well below the melting point of copper.

MD simulation of tensile behaviour has been carried out at 10 K with a strain rate of $1\times10^9$ s$^{-1}$ and the results obtained are shown in Fig. 2. It can be seen that the tensile behaviour in both the thermal loading conditions was characterized by sharp rise in stress up to onset of yielding followed by significant stress drop. The yield strength value obtained was about 9.7 GPa at a strain level of 0.1. With further straining after stress drop, the stress increased marginally followed by gradual drop in stress until failure. The observed behaviour is consistent with those reported for single crystal Cu by several investigators [1,6,7]. In addition, it has been seen that the tensile stress values in plastic regime were marginally higher for isothermal condition than that obtained under adiabatic condition (Fig. 2). In adiabatic deformation, the temperature rise in plastic deformation regime lowers the flow stress and increases the ductility as compared to isothermal conditions (Fig. 2).

An important finding in the present investigation is that the stress response in plastic regime for both the thermal loading conditions can be correlated with the evolution of dislocation density. Dislocation density as a function of strain was estimated and superimposed on the stress-strain curve for adiabatic conditions in Fig. 3. It can be seen that the overall stress response could be inversely correlated to dislocation density. Similar behaviour was seen for isothermal condition except that the dislocation density was marginally higher at any given value of strain. The fluctuations in stress-strain curve can also be correlated with the variation in dislocation density due to continuous generation and escape of dislocations to free surfaces.

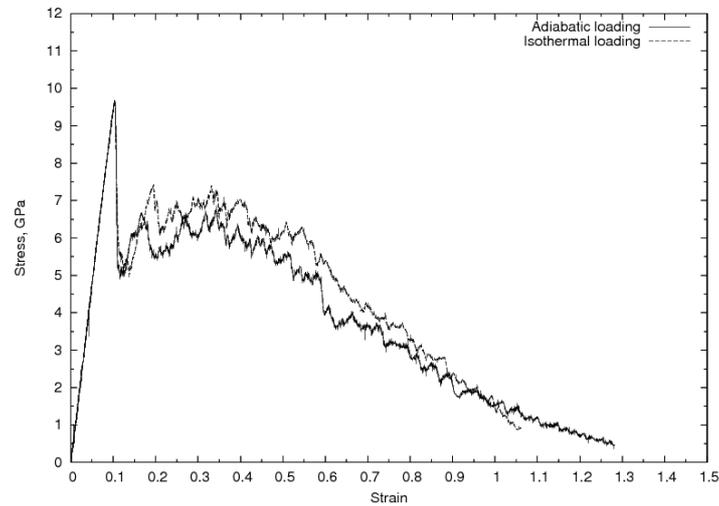

Fig. 2. Tensile stress-strain curves for single crystal copper under isothermal and adiabatic conditions.

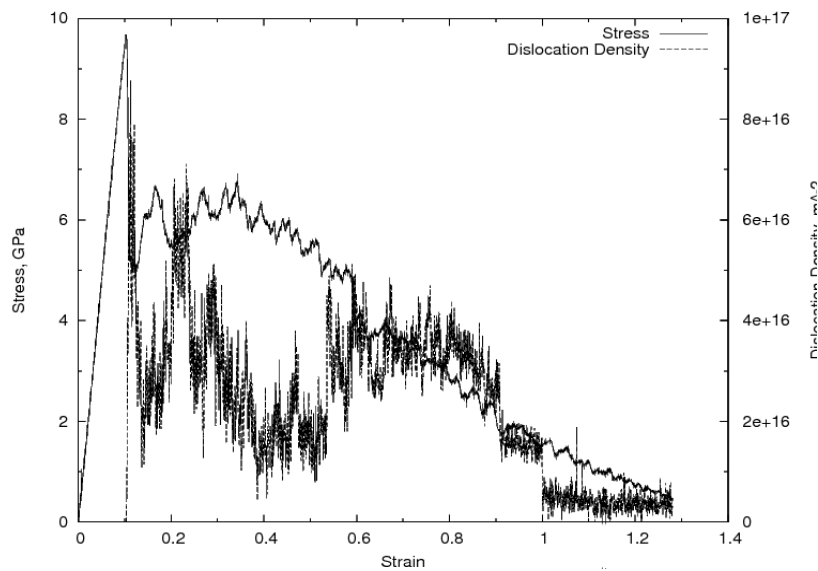

Fig. 3. Tensile stress-strain response under adiabatic conditions superimposed with dislocation density.

The micro-structural behaviour of copper during tensile deformation is depicted in Figs. 4 and 5 which can be correlated to tensile behaviour. Drastic drop in stress value observed immediately after yield strength is attributed to dislocation generation in an initially defect free single crystal copper (Figs. 4a-b). Nanoscale materials are unlikely to activate Frank-Read sources because of their reduced dimensions. It is therefore assumed that plasticity can then be initiated by the nucleation of dislocations from regions such as crack tips, surfaces, interfaces or grain boundaries. It is expected that the dislocations preferably nucleate at stress concentrating sites on the surface in absence of other potential nucleation sites. The observed nucleation of dislocations at sharp corners on the surface of rectangular specimens (Fig. 4a-b) confirms this view. Typical dislocation structure observed during plastic deformation of Cu is shown in Fig. 4d.

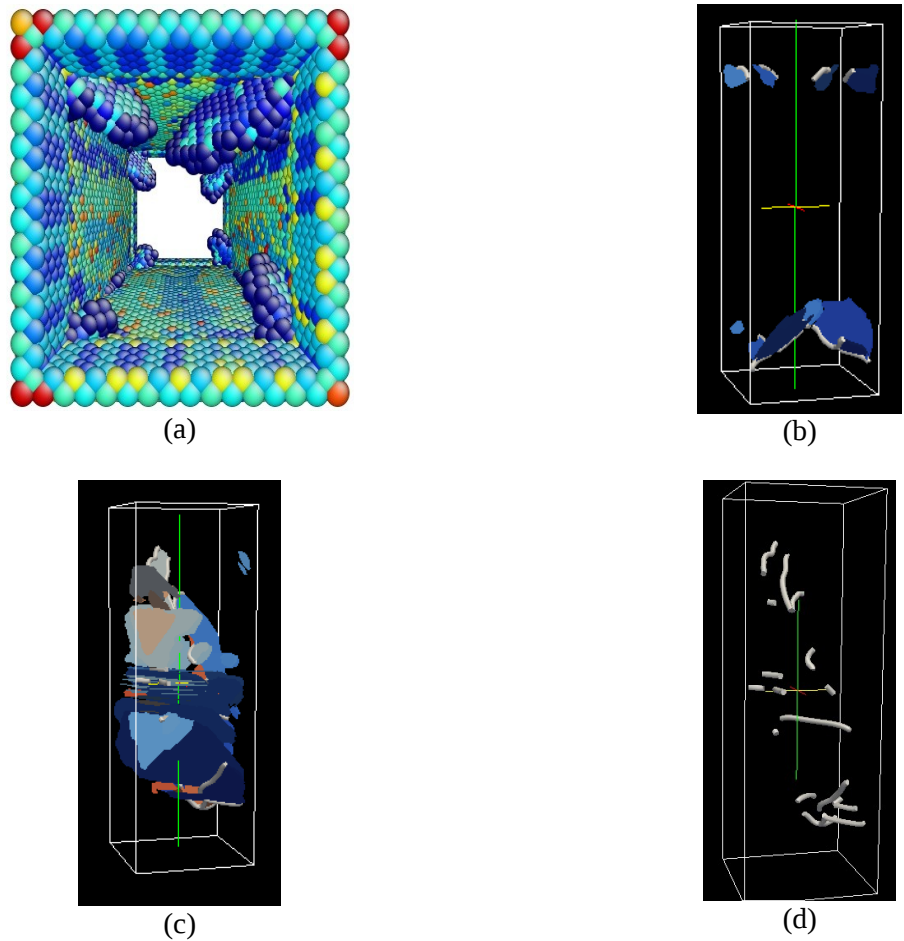

Fig. 4. (a) Dislocation initiation at corners as observed using Atom-Eye in axial direction (b) dislocation nucleation (c) stacking faults along with dislocations and (d) dislocations alone at relatively higher strain of 0.19. Figures (b-d) were DXA snapshots.

Figure 5a demonstrate typical deformation pattern in the specimen at the strain level of 0.19. In addition to deformation by slip, significant lattice reorientations are seen in the specimen (Fig. 5a), which can influence the plastic deformation. Figures 5b and 5c represent the necking behaviour under adiabatic and isothermal conditions. It can be seen that post necking deformation is observed to be more under adiabatic condition (Fig. 5b) than that obtained under isothermal condition (Fig. 5c). In addition to these observations, several other important features affecting the deformation behaviour observed in the present investigation are described in the following sections.

**3.2 Stacking faults during tensile deformation**

Shockley partials are observed to initiate from surface at the onset of yielding and propagate into the specimen on further straining (Fig. 4a). In addition to this, stacking faults were seen during yielding (Fig. 4b) and also at all stages of deformation at higher strains (Fig. 4c) by using DXA. Similar observations were made by Kolluri et al [7] and Hirel et al [8]. Dislocations in nanocrystalline materials have been shown to emerge from the surface as partial dislocations loops and propagate to the opposite surface to create a step [8]. These Shockley partials dragging a stacking fault which can propagate only when they exceed a critical radius. This mechanism requires thermal activation [8]. Stacking faults, which are abundantly present in FCC metals, play a significant role in the dissociation, cross-slip, and eventual annihilation of dislocations in small-volume structures of FCC metals thereby influencing the plastic deformation [7].

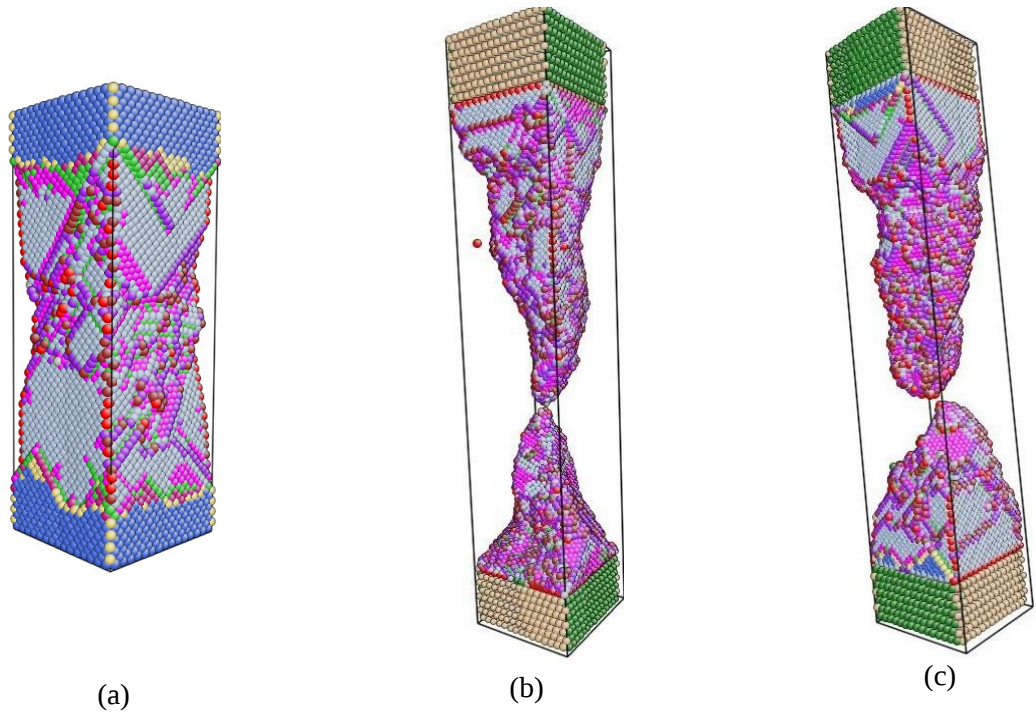

(a)    (b)    (c)

Fig. 5. The atomic snapshots showing the tensile deformation behaviour of Cu (a) Deformation in the stress drop region following the yielding at a strain 0.19, (b) necking under adiabatic condition and (c) necking under isothermal condition.

**3.3 Phase transformation**

Partial phase transformation in Cu from fcc to bcc structure has been observed during deformation. Immediately after yielding, about 0.16% of Cu atoms in the bulk transformed to bcc structure and the rest of the bulk remained as fcc structure. This transformation has been determined by common neighbour analysis. Phase transformation during tensile deformation of single crystal nanowires is not uncommon. Investigation on the tensile deformation of single crystal bcc iron indicate that just after the yielding, the single crystal bcc iron almost completely transformed to fcc resulting in softening in the stress-strain response. It has been shown that during straining, several metals including gold nanowires can transform from fcc to bct structure due to surface-stress-effects [9]. It has been suggested that the phase transformation is controlled by wire size, initial orientation, boundary conditions, temperature and initial cross sectional shape. [9]. The low amount of phase transformation observed in the present study in fcc Cu can be ascribed to the effects associated with unfavorable loading conditions.

**3.4 Pentagonal structure in necking region**

The MD simulation of tensile behaviour of Cu nanowire showed that the material invariably fails by necking phenomenon (Figs. 5c-d). An interesting observation is that pentagonal arrangement of Cu atoms forms in the necking region (Fig. 6). In Cu nanowires, atomistic simulations under various conditions have shown to produce different polygonal cross-sections [10]. Pentagonal structure observed in single crystal Cu nanowire during necking has been analyzed by Sutrakar and Mahapatra [5]. They have shown that pentagonal nanostructure is a stable structure and assist in extensive plastic deformation compared to structures without pentagons. Large surface stresses [9] and lateral lattice orientations [11] are shown to be responsible for pentagonal structure formation.

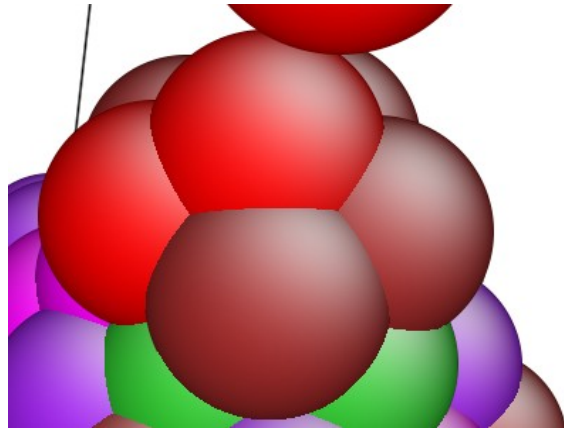

Fig. 6. Pentagonal structure as near necking point.

## 4. Conclusions

(i) The tensile stress-strain response was characterized by steep rise in stress value up to yielding followed by rapid drop in stress due to nucleation of dislocations.
(ii) The flow stress was lower and strain to fracture was higher under adiabatic conditions as compared to isothermal conditions.
(iii) Dislocation density as a function of strain is inversely related to stress response.
(iv) Several micro-structural features including stacking faults, phase transformation and pentagonal structure were observed during plastic deformation.